# Voltage enhancement and loss minimization in a radial network through optimal capacitor sizing and placement based on Crow Search Algorithm

## Stephen W. Mathenge*[1], Edwell. T. Mharakurwa[1], Lucas Mogaka[2]

[1]Department of Electrical and Electronic Engineering, Dedan Kimathi University of Technology, Private Bag 10143, Nyeri, Kenya

[2]Department of Electrical and Electronic Engineering, Meru University of Science and Technology, P.O BOX, 972-60200, Meru, Kenya

*Email: stephen.wamae@dkut.ac.ke

*Abstract*

*In power systems, the distribution network is pivotal for consumers because it is the final stage in the delivery of electricity from the generation plants to the end users. Reactive power demand from consumers can result in challenges like low power factor, diminished voltage, and heightened power losses. The primary challenge encountered when utilizing a radial system as a distribution network is the voltage drop, which leads to distortion in the voltage profile across the entire network. This research focuses on optimizing network performance by appropriately sizing and placing capacitors based on the Crow Search Algorithm (CSA). The results from this approach were compared with Particle Swarm Optimization (PSO), Artificial bee colony, Cultural Algorithm, Firefly Algorithm, Genetic Algorithm, Invasive Weed Optimization, and Teacher Learner Based Optimization methods illustrated in the study where CSA's effectiveness, demonstrated a notable 30.41% and 29.33% overall reduction in active and reactive power losses compared to the base case. The lowest bus voltage, notably at bus 18 which happens to be the longest in the IEEE 33 Radial Distribution Network (RDN) topology, improved from 0.8820 in the base case to 0.908 with CSA, showcasing a 32.9% enhancement in voltage deviation. This optimization led to a significant 3.841pu reduction in capacitor cost, highlighting CSA's efficacy in enhancing RDN performance and cost efficiency.*

## 1. Introduction
The demand for energy, particularly in developing countries continues to rise steadily due to factors such as industrialization and population growth. Consequently, the management of losses in transmission and distribution networks becomes paramount for achieving optimal performance, as emphasized in [1, 2]. These losses often stem from stressed lines, which occur when electrical distribution lines are operating near or beyond their maximum capacity.

Stress on power lines can be due to high demand, inadequate infrastructure, or poor maintenance. When lines are stressed, several negative implications arise beyond just increased losses such as increased heat, voltage drops, equipment overload, power quality issues, increased maintenance costs, safety risks, reliability and stability of the system as indicated in [38-39]. Recognizing energy efficiency and conservation as pivotal factors in industrial, commercial, and institutional facilities are essential for mitigating these issues then implementing measures such as: demand-side management, upgrading infrastructure, preventive maintenance and use of smart grid technologies can enhance their performance, reduce energy losses, and improve overall reliability and safety of the power distribution network as demonstrated in reference [27-29].

However, there are several other general methods that can be employed to improve system reliability and efficiency such as, load balancing, upgrading infrastructure, preventive maintenance, demand-side management, automation, smart grids, energy storage systems, distributed generation, power factor correction, network reconfiguration, and voltage optimization as detailed in reference [27-30]. After considering these general methods, the strategic placement of capacitors becomes a specific and effective technique for enhancing network reliability and efficiency. Capacitors help in, voltage regulation, loss reduction, complementing the broader strategies and enhancing overall network performance as demonstrated in reference [34-35].

Without diligent monitoring and regulation, the escalating demand for power can result in significant power interruptions. In [1], it was observed that the power utility encountered overall losses which exceeded their recommended threshold of 19.9% set by the regulatory authority, reaching 22.43%. This highlights the critical need for effective measures to manage and optimize the power distribution network. However, most of the existing research primarily focuses on distributed generation (DG) integration, leaving a gap in addressing the sustainability of protective devices within the system. Despite the benefits of DGs, their introduction increases network current, potentially leading to frequent tripping of protective devices unless they are modified.

A significant body of research has been dedicated to improving the efficiency and reliability of radial distribution networks through various optimization techniques. These methods primarily aim to minimize power losses, enhance voltage profiles, and ensure the stability of the network under varying load conditions.

### 1.1 Related work on Optimization techniques applied in Radial Network

In recent years, researchers have explored various AI, analytical [3-6], and metaheuristic methodologies [7-9] for optimization problems, particularly for power loss reduction. The Crow Search Algorithm (CSA) has emerged as a versatile tool, finding applications in various areas, including Combined Economic and Emission Dispatch (CEED) [10], hybrid methods that combine CSA with the Grey Wolf Optimizer and the implementation based on Dynamic Fuzzy Learning Strategy for large-scale optimization [11]. CSA's effectiveness is further demonstrated in Binary Crow Search Algorithm applications, such as the Two-dimensional Bin Packing Problem [12], and in minimizing generation costs in CEED scenarios [13].

Several optimization algorithms have been proposed to address power loss minimization, though not all consider reactive power loss explicitly. Notable methods include the Salp Swarm and Gray Wolf Optimizer [14], Moth Flame Optimization [15], and Gray Wolf Optimization [16]. Harmony Search [17], Elephant Herding Optimization [18], and modified Selective Particle Swarm

Optimization [19] are also used. The Hybrid Firefly and Particle Swarm Optimization [20] focuses on generation cost minimization and power loss reduction, while Fractional Order Particle Swarm Optimization [21] is tailored for reactive power optimization. Other strategies like Improved Transient Search [22], Antlion Optimizer (ALO) [23], Index Vector Method, and Modified Whale Optimization Algorithm [24] target network reconfiguration to minimize losses. These diverse methodologies demonstrate the multifaceted approaches researchers use to tackle power loss and reactive power optimization in power systems.

Reference [25] highlights that each optimization algorithm offers distinct advantages in efficiency and solution quality but also faces implementation and tuning challenges. The improved metaheuristic method in [26] optimizes network reconfiguration and DG allocation but has practicality challenges. The hybrid Whale Optimization (WOA) and Salp Swarm (SSA) algorithms (WOA-SSA) in [27] effectively places and sizes multiple DGs to reduce power losses, though it is complex and demanding to implement. The PSO-OS algorithm in [28] enhances solution quality and convergence speed for DGs and Shunt Capacitors (SCs) placement, but requires careful parameter tuning and integration. The index-based method in [29] is simple and efficient for optimal DG placement, but faces challenges related to data accuracy, flexibility, and scalability.

Recent studies have explored various optimization strategies for power distribution networks. Authors in [30] provided an overview of methods for integrating Distributed Generations (DGs) and Capacitor Banks (CBs), highlighting both benefits and implementation challenges. The Multiverse Optimizer for shunt capacitor placement, detailed in [31], shows advantages in solution quality, speed, and robustness, but faces challenges in complexity, parameter sensitivity, and computational demands. Similarly, reference [32] discusses distribution network reconfiguration optimization, noting benefits in solution quality and speed but limitations come in form of complexity and scalability. Reference [33] introduced a multi-objective Optimal Power Flow (OPF) method that improves voltage stability and power system performance but requires careful management of problem formulation, computational demands, and data requirements for practical application. Furthermore, specific challenges such as optimal placement and sizing of capacitors have been addressed, as evidenced in [34, 35]. This diverse range of optimization techniques underscores the ongoing efforts to enhance the efficiency and reliability of power systems through advanced algorithmic approaches.

It is noted in literature that work has been done in trying to minimize power losses through implementation of diverse algorithms, however, thorough discussions have not been exhaustedly on adapting protective devices for higher network currents, posing safety and efficiency risks. To address this, there should be a collaborative effort to explore new technologies, update standards, conduct risk assessments, and promote industry dialogue. This would improve the resilience and safety of electrical infrastructure. Incorporating well sized capacitors in certain regions of the power system can enhance voltage profiles and reduce stress on protective devices.

The goal of this paper was to optimize the performance of a RDN by effectively integrating capacitors based on the Crow Search Algorithm (CSA). The implemented approach aimed to minimize power losses and improve voltage profiles, addressing the increasing demand for efficient and reliable power distribution. The modelled network outcome was achieved through simulations where Backward/Forward Sweep (BFS) load flow method was implemented in order to achieve the result from the RDN. The backward/forward sweep method was preferred method because the outcome obtained compared to the Newton Raphson, Gauss Seidel or Fast Decoupled methods. Where the Newton-Raphson method offers high accuracy, it often requires the highest computational time, especially for large-scale systems. The Gauss-Seidel method may be slower due to its sequential updating approach, while the Fast Decoupled method tends to be faster, particularly for large-scale systems with weak coupling between variables. The radial distribution has very high values of R/X ratio in the network. Determining power losses at individual buses and assessing voltage profiles at each node becomes easier when utilizing the backward/forward sweep method.

In order to improve system stability and reliability an IEEE 33 bus system model in Matlab was modelled where, a baseline for voltage profiles and power losses was established through load flow analysis without capacitors. The CSA's performance was compared to other metaheuristic algorithms, in addressing power reliability concerns.

This paper has been divided into sections where section 2 outlines the mathematical problem formulation of the RDN, with section 3 describing the proposed CSA. Section 4 describes the results achieved from the CSA and the base case. Section 5 demonstrates the effectiveness of the CSA compared with PSO and other optimization techniques in the RDN while section 6 highlights the conclusions and recommendations.

## 2. Mathematical problem formulation

The primary objective function of this study is to minimize the active and reactive power losses in the RDN while ensuring all the constraints are within the allowable ranges.

### a) Objective function

Backward and forward sweep approach was used as a power load flow tool where the objective was to reduce the power losses and the costs involved on the capacitors in the overall network. The main aim of minimizing the losses through capacitor sizing and placement is by minimization of loss function. Equation (1), illustrates the active power and equation (2), reactive power losses that were used to achieve the minimization [40, 41].

$$minimize(x) = \sum_{i=1}^{nb} P_{cost}.P_{i\,loss} + \sum_{j=1}^{nc} cap_{i\,cost}.cap_j + \sum_{k=1}^{nb} |V_{k-1}| \quad (1)$$

$$minimize(x) = \sum_{i=1}^{nb} Q_{cost}.Q_{i\,loss} + \sum_{j=1}^{nc} cap_{i\,cost}.cap_j + \sum_{k=1}^{nb} |V_{k-1}| \quad (2)$$

where:
$x$ = best sizing and capacitor placement, $nb$ = Number of buses, $P_{cost}$ = real power loss, $P_{i\,loss}$ = power loss in bus i, $Q_{cost}$ = real power loss, $Q_{i\,loss}$ = power loss in bus i, $nc$ = Number of capacitors, $cap_{j\,cost}$ = cost of capacitor j, $cap_j$ = size of capacitor j, $V_k$ = voltage magnitude of bus k.

Equation (3), illustrates the total active power losses, equation (4), illustrates the total reactive power losses while equation (5), highlights the sizing formulas in the entire RDN [38 - 40].

$$Pi\ loss = \sum_{j=1}^{nb}\left(\frac{P_{ij}^2 + Q_{ij}^2}{V_i^2}\right) * R_{ij} \qquad (3)$$

$$Qi\ loss = \sum_{j=1}^{nb}\left(\frac{P_{ij}^2 + Q_{ij}^2}{V_i^2}\right) * X_{ij} \qquad (4)$$

where:
$Pi$ loss = power loss in bus $i$, $Qi$ loss = power loss in bus $i$, $nb$ = Number of buses, $P_{ij}$ = active power at the sending end (i) and the receiving end (j), $Q_{ij}$ = active power at the sending end (i) and the receiving end (j), $V_i$ = voltage magnitude at the sending end, $R_{ij}$ = resistance at between node $i$ and node $j$, $X_{ij}$ = resistance at between node $i$ and node $j$

b) Constraint
i. Voltage limit for capacitor placement and sizing with per unit limits of $0.9 \leq V_i \leq 1$
where $V_i$ = voltage size in bus $i$

ii. Capacitor size must be less than total effective reactive power

$$\sum_{i=1}^{nc} cap_i < \sum_{j=1}^{nb} Q_j \qquad (5)$$

where:
$nc$ = number of capacitors, $cap_i$ = size of capacitor i, $nb$ = number of buses, $Q_j$ = reactive power KVAR on bus i

iii. Optimal location of the capacitor between buses with the limit $2 \leq L_i \leq L_{max}$
where: $L_i$ = location of the capacitor in bus i, $L_{max}$ = max bus loation

## 3. Methodology

In this section, an approach to optimizing the RDN based on the CSA is presented. The CSA, inspired by the intelligent foraging behavior of crows, has shown significant potential in solving complex optimization problems. The methodology employed focuses on optimizing capacitor sizing and placement within the RDN to enhance system stability and efficiency. The optimization process employs two critical parameters of the CSA: flight length (fl) and awareness probability (Ap). These parameters are integral to the algorithm's search dynamics. The flight length determines the step size taken by the crows during the search process, influencing the exploration of the solution space. Meanwhile, the awareness probability dictates the likelihood of a crow choosing to follow its memory or randomly search for a better solution, thereby balancing exploration and exploitation.

To demonstrate the efficacy of CSAA approach, an IEEE 33 bus, 11 kV RDN as shown in Figure 1, was modelled in Matlab. Initially, load flow analysis was conducted without capacitors to establish a baseline for voltage profiles and power losses. The model was then subjected to the CSA-based optimization process, where various configurations of capacitors were evaluated to achieve optimal performance. This approach suggests that advanced methods like the CSA can offer significant improvements, though ongoing research and development are necessary to fully address the complexities of the RDN.

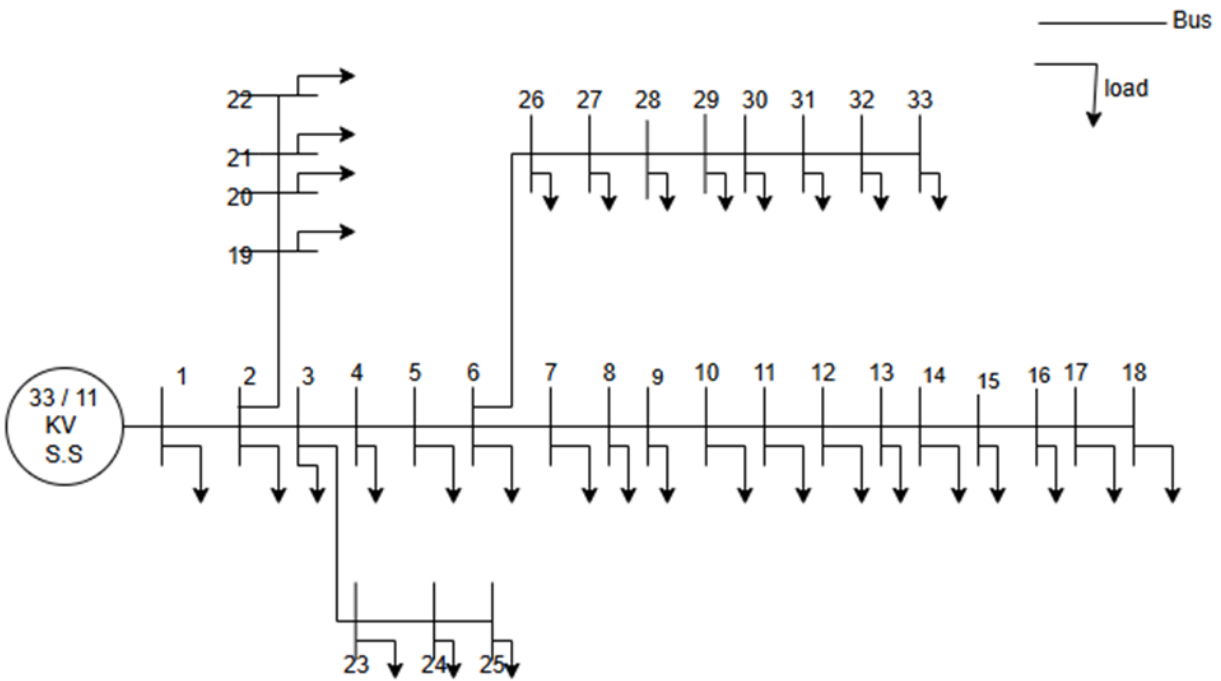

Figure 1: Single line diagram for standard IEEE 33 Bus RDN, 33/11kV substation

### 3.1 Crow Search Algorithm

The CSA is population-based metaheuristic optimization algorithm that was proposed by Askarzadeh, [37]. The formulation of the CSA in this paper was aimed at delivering a precise, efficient, and adaptive method to significantly improve the performance and reliability of power distribution networks. The algorithm balances between exploitation and exploration processes with the help of two parameters, the flight length (fl) and the awareness probability (Ap). To implement the CSA, the steps shown in Figure 2 and the flow chart in Figure 3 were followed. These figures provide a detailed guide on the execution of the algorithm.

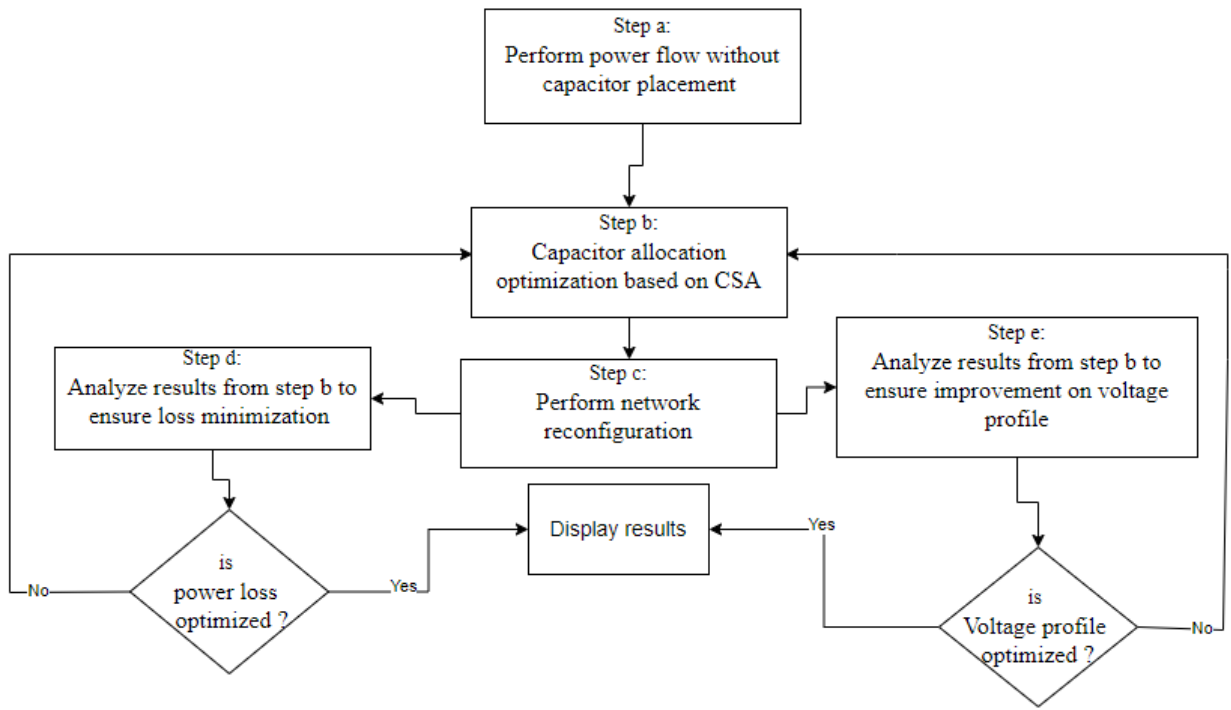

Figure 2: Block diagram of the generalized CSA, optimal capacitor sizing and placement allocation based Ap and fl

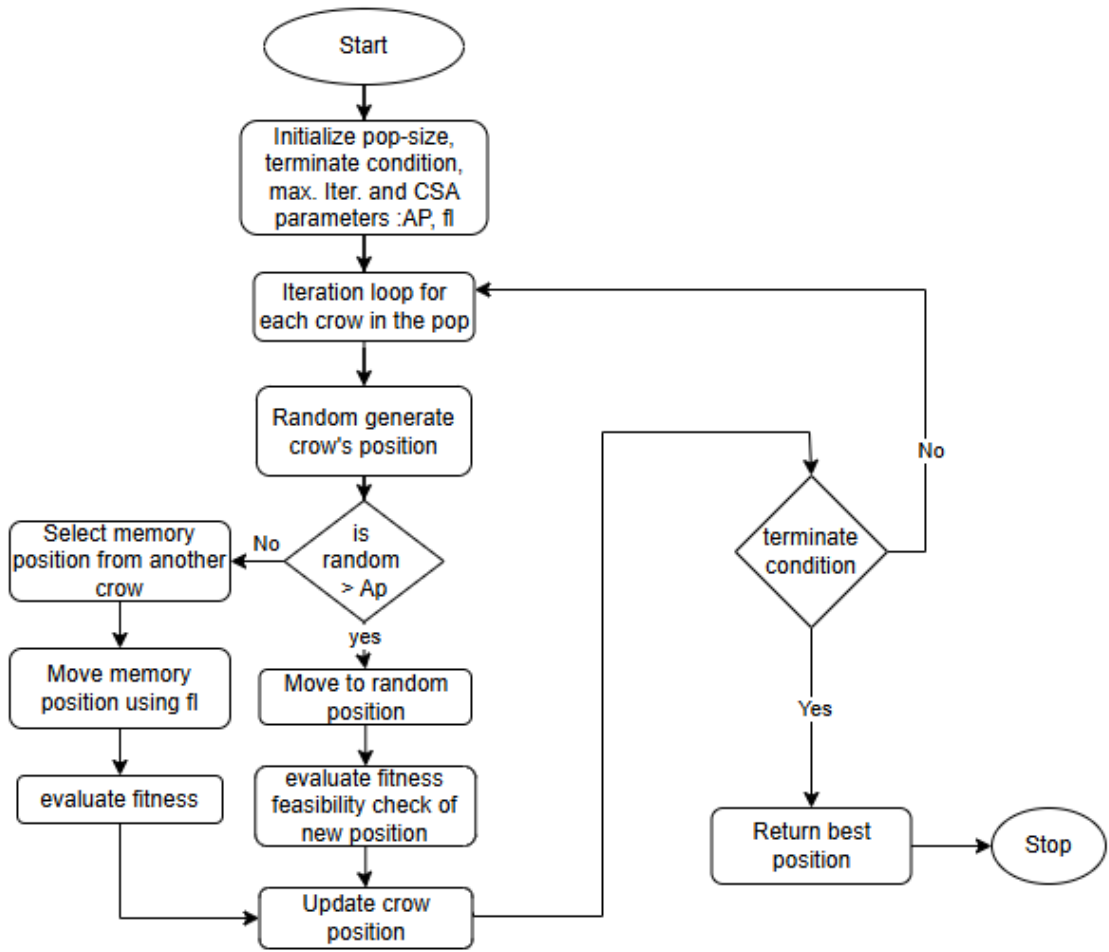

Figure 3: Step by step process of sizing and placement of capacitors based the CSA

### 3.2 Implementation of the CSA

Load flow analysis is a fundamental tool in power system analysis. It enables engineers to compute the voltages and currents throughout the network, ensuring that the system operates efficiently, reliably, and within safe limits. By providing insights into the power flows and losses, load flow analysis supports network planning, operational decision-making, and stability assessments, making it indispensable for modern power system management.

The load flow implemented in this study is the BFS which involves two processes at a time. The backward sweep utilizes either current or power flow solutions from the receiving end branch node toward the sending end node and the forward sweep utilizes the voltage calculations starting from the sending end node to the receiving end node. At the time the backward sweep is being executed the voltage is held constant and the same happens in the forward sweep where the current or power is held constant. The convergence is tested after each and every $n^{th}$ iteration.

The backward sweep method is illustrated by equation (6), which indicated how the current was calculated [38 - 40]:

$$I_i^k = \frac{P_i + Q_i}{V_i^{k-1}} \quad i = 1,2 \ldots n \tag{6}$$

where:

$I_i^k$ is the current injected at node i in the k$^{th}$ iteration, $P_i + Q_i$ is the apparent power injected at node i, $V_i^{k-1}$ is the voltage at node $i$ at the k$^{th}$-1 iteration

While performing the backward sweep the beginning node current $I$ and the end is node current $J$ in that case the equation (7), is used. For the forward sweep method where current is held constant equation (8) is utilized, [39, 40]:

$$J_{i-1,i}^k = I_i^k + \sum(j_{i,i+1}^k) \quad i = 1,2 \ldots \ldots n \tag{7}$$

$$V_i^k = V_{i-1}^k - J_{i-1}^k Z_{i-1,i} \quad i = 1,2 \ldots \ldots n \tag{8}$$

where:

$J_{i-1,i}^k$ is current at node i towards node i-1, $\sum(j_{i,i+1}^k)$ is the sum of all the branch currents from node i, $V_i^k$ is the voltage at node i at the k$^{th}$ iteration, $V_{i-1}^k$ is the immediate node to node i, $J_{i-1}^k$ is the branch current node i and the immediate node, $Z_{i-1,i}$ is the branch impedance from node i going forward.

For each node at every level, the voltage drop caused by the parent branch of a certain node is calculated and subtracted from its parent node's voltage, updating the solution set. This iterative process of backward and forward sweeps continues until convergence criteria are satisfied. The flowchart depicting the Backward/forward sweep is shown in Figure 4.

It is essential to emphasize that, unlike many other methods such as Newton Raphson, Gauss Seidel, or Fast Decoupled, Backward/Forward Sweep implementation does not require the network to be classified or partitioned into main and derived lines, nor does it involve matrix multiplication. Solutions are obtained simply by adding and subtracting both the voltages and currents in the network. Additionally, this method is node-centric and does not depend on node numbering or naming conventions, as nodes interact through parent-child relationships. Once a node knows its parent or children, it can automatically retrieve their names and addresses in the solution set.

The flowchart in Figure 4 illustrates the CSA integrated with the BFS power flow method. This detailed process ensures that the CSA effectively determines the optimal sizes and placements of capacitors within the radial distribution network, improving system performance without requiring node renumbering.

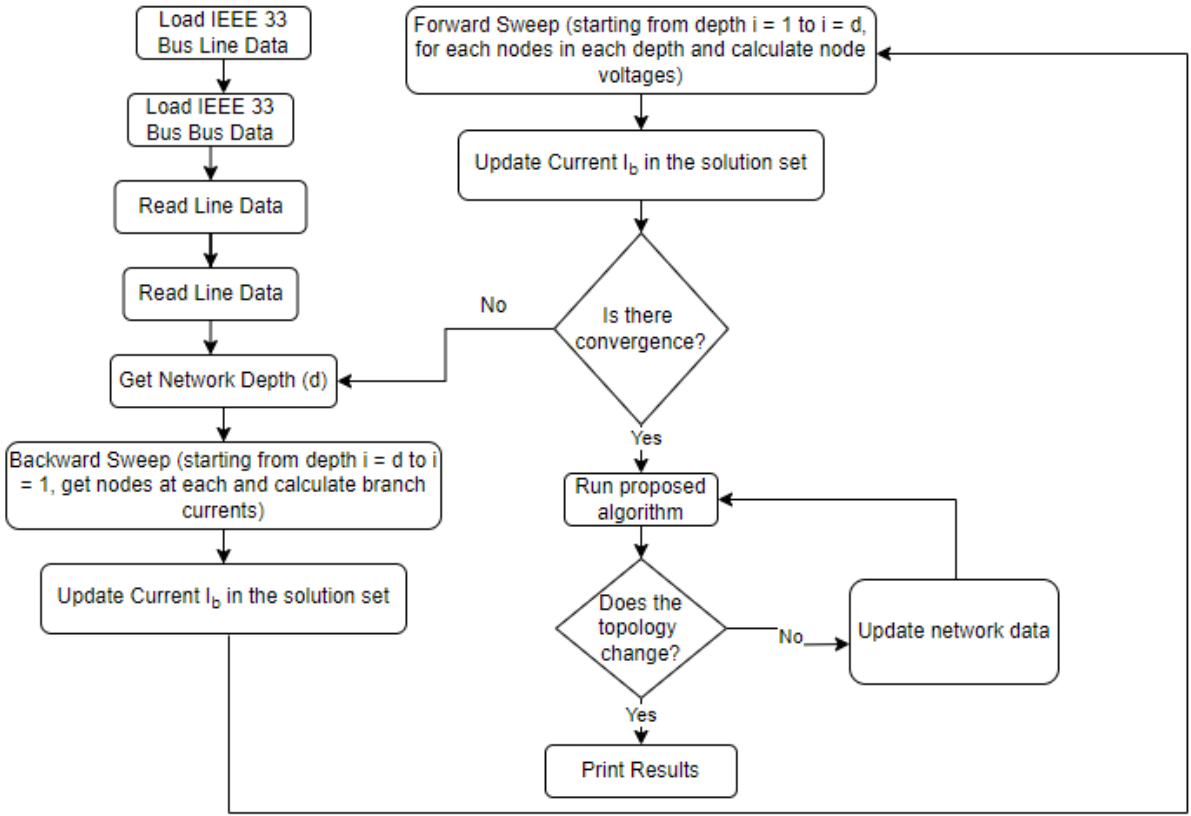

Figure 4: Flow chart illustrating the implementation of Backward/Forward Sweep and CSA

The following are the evaluation procedures for Base Case and CSA Case that was implemented in this study:

### a. Performing Base Case Evaluation

The network modeling was performed by creating a standard IEEE 33 bus, 11 kV for a RDN in MATLAB. Which included 33 buses, 32 lines, and loads at each bus. After the modelling was performed then Load Flow Analysis without Capacitors was performed using the BFS method in order to compute the voltage profile, active power losses, and reactive power losses across the network. This was followed by documenting the Base Case Results through recording the voltage levels at each bus and calculate the total active and reactive power losses in the network. The final stage was identifying any bus with voltage violations (i.e., voltages outside the acceptable range).

### b. CSA Implementation for Capacitor Placement

The CSA was initialized by first defining its parameters, including the number of crows (population size), flight length (fl), awareness probability (Ap), and the number of iterations. Then randomly initialized the positions of crows representing potential capacitor sizes and locations within the network. Then the Load Flow Analysis with Capacitors was to be ran so as to determine each crow's position (i.e., each potential solution of capacitor sizes and placements), using the BFS method. Then evaluating the fitness of each solution based on the reduction in active and reactive power losses and improvement in voltage profiles. However the optimization process which involves the use of the CSA to iteratively update the positions of the crows, guided by their memories and random exploration influenced by fl and Ap. After each update, load flow analysis was performed and the evaluation of the new fitness of the proposed solution. This was followed by updating the crow's memory depending on the new position results in better fitness. Convergence followed as the final step where continue iterative process was performed until the stopping criterion (such as the maximum number of iterations or convergence to an optimal solution) is met. Upon where recording of the best solution found, representing the optimal capacitor sizes and placements followed as the final step.

### c. Comparative Analysis

Evaluation of CSA case was done by implementing the optimal capacitor sizes and placements found by the CSA in the RDN model. This was done by performing a load flow analysis using the BFS method with the capacitors in place by computing the voltage profile, active power losses, and reactive power losses in the network.

### d. Comparison with Base Case

Compare the results of the base case (without capacitors) and the CSA case (with optimal capacitors) was to follow. This was followed by highlighting the improvements in voltage profiles, and reductions in active and reactive power losses.

### e. Documenting Results

The results from the voltage levels at each bus for both the base case and CSA case were documented which were followed by a summary of the total active and reactive power losses for both scenarios. Discuss the effectiveness of the CSA in enhancing network performance was the last step.

The procedures outlining the methodology used in the study was outlined in the steps below:

### i. Base Case Evaluation

Description of network modeling in MATLAB was done followed by steps for performing load flow analysis using BFS without capacitors and then documentation of base case results (voltage profile, power losses).

### ii. Crow Search Algorithm Implementation

Initialization of CSA parameters and crow positions after which the load flow analysis with proposed capacitor placements was done iteratively by optimization process based on the CSA and finally the convergence criteria and identification of optimal solution.

### iii. Comparative Analysis

Load flow analysis with optimal capacitors in place was done where comparison of voltage profiles and power losses between base case and CSA case and finally summarizing the improvements and effectiveness of CSA. The base case and the CSA case were evaluated and the results showed remarkable improvement. This was followed by data verification through Particle Swarm Optimization and other optimization algorithm outlined in section 5.

## 4. Simulation Results

This section presents the outcomes of the simulations conducted to evaluate the performance of the CSA against the base case in enhancing the voltage profile and reducing network losses in a radial distribution network. The results are divided into the base case study results and the optimized case results, providing a comparative analysis of the network's performance before and after the application of the optimization techniques.

### 4.1 Base Case Study Results

Before capacitor sizing and placement (Base Case), the study observed an initial active power loss of 281.58 kW and reactive power of 187.96 kVAr across the entire network. Notably, the maximum voltage recorded was 1 p.u. on bus 1, while the minimum voltage, at 0.8820 p.u., was noted on bus 18. The lower voltage on bus 18 was attributed to the length of the line, which from Figure 1 and taking a per unit length from one bus to the next, being the longest on an 11 kV base voltage. Figure 5 below provides a visual representation of the voltage profile of the Base Case. By taking bus 6, 18 and 33 for illustration purposes, it is observed in Figure 5 that bus 6 was within the set limits while bus 18 and 33 are below the set limits of $0.9 \leq V_i \leq 1$ and hence the need to enhance the voltage profile.

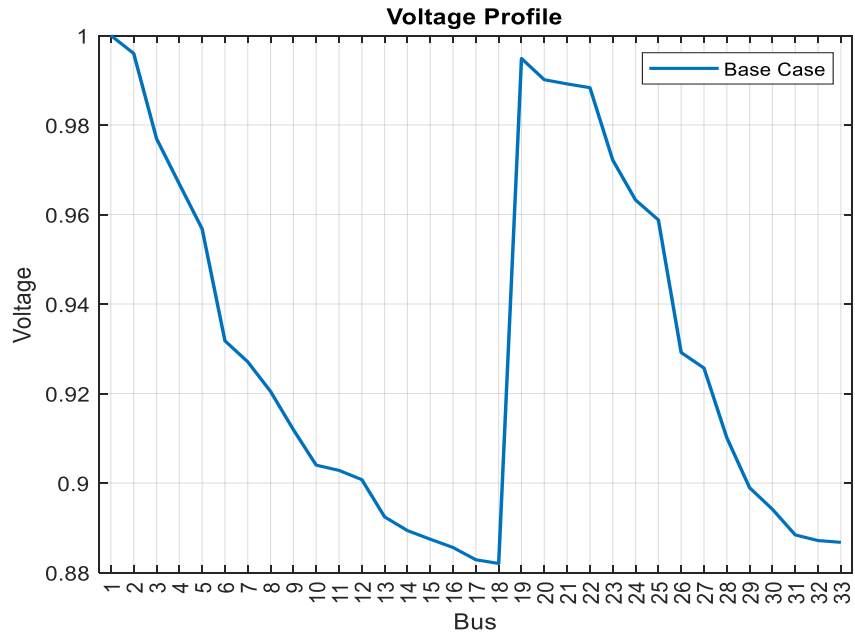

Figure 5: Voltage profile for base case

By changing the topology of a distribution network, it requires careful analysis and planning to manage its impact on voltage profile and power. Adjustments in the network design, additional equipment, and advanced control mechanisms are often necessary to mitigate negative effects and ensure efficient and reliable operation. While RDNs may not match the redundancy and reliability of ring and interconnected systems, they provide a cost effective, simple and reliable solution for many applications, particularly in small or less densely populated areas where the load is relatively predictable and manageable. These advantages make the RDNs more popular choice for many utility companies, especially in regions where cost and simplicity are critical factors. Hence the choice of this topology.

Figure 6 shows the active power losses for each bus, indicating that bus 2 recorded the highest loss of 71.3928 kW, followed by bus 5 with 53.2985 kW for the base case. This distribution reveals that certain buses, notably bus 2 and bus 5, are experiencing significant losses in active power. Such losses may arise due to factors like resistive losses in distribution lines. Addressing these losses is crucial for enhancing the overall efficiency and reliability of the power system, potentially through measures such as capacitor sizing and placement, or optimization of operational strategies.

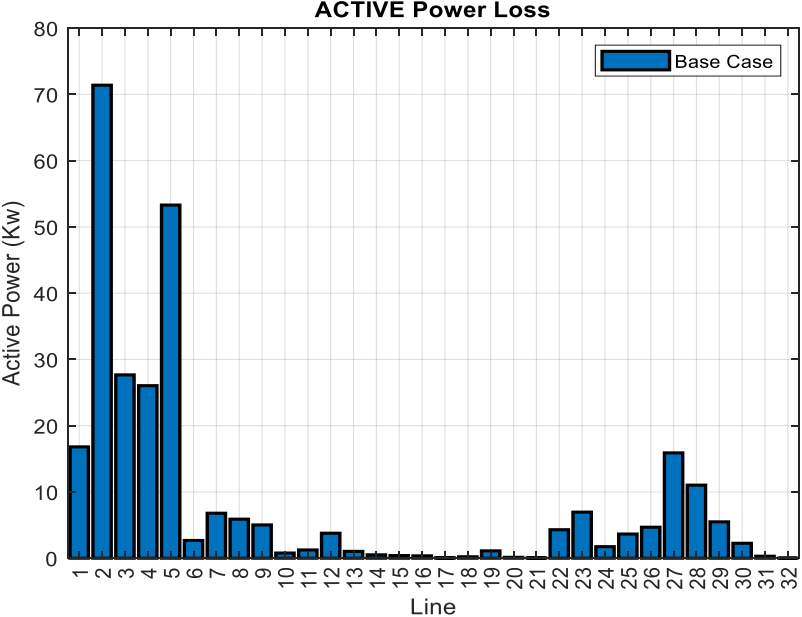

Figure 6: Base case active power loss

Figure 7 provides a visualization of the reactive power distribution for the base case scenario where the reactive power losses for each bus are illustrated, with bus 5 exhibiting the highest loss of 46.0098 Kvar, followed by bus 2 with 36.3625 Kvar. This distribution suggests that certain buses, particularly bus 5 and bus 2, are experiencing notable losses in reactive power. These losses could be attributed to various factors such as impedance mismatches, voltage drops, or inefficiencies in the network configuration, highlighting areas that may require optimization or corrective measures to improve system performance.

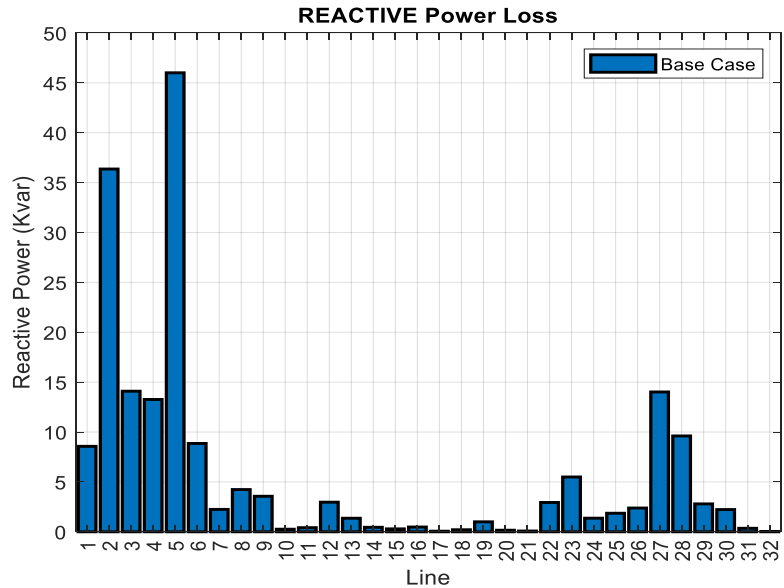

Figure 7: Base case reactive power loss

## 4.2 Crow Search Algorithm

The CSA was applied in order to optimally size and place the capacitor in the network, where the voltage profile saw an enhancement from 0.8820 p.u to 0.90798 p.u on bus 18. Figure 8 displays a comparison of voltage profiles between CSA and the base case, highlighting a positive influence on voltage profiles across all buses. This enhancement stems from a reduction in current supply from the source to the load, consequently leading to decreased power losses in each branch. The optimization process helps in maintaining voltage levels within desired limits across the network as evidenced from the application of the CSA.

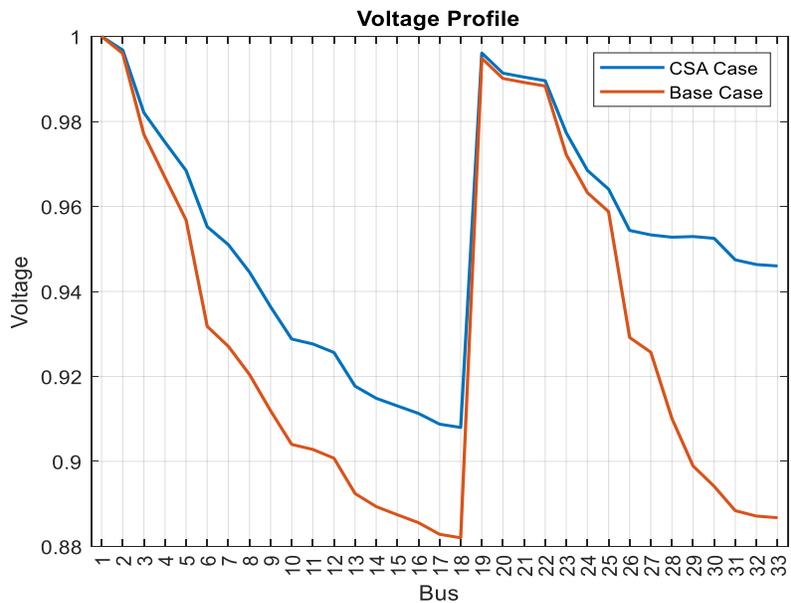

Figure 8: Comparison of voltage profiles when capacitors are optimally placed using CSA and base case

The CSA demonstrated a notable enhancement in active power loss, as evidenced by Figure 9. The CSA is designed to perform an efficient search, balancing exploration and exploitation to navigate the solution space effectively. This capability enables the algorithm to identify optimal or near-optimal solutions for minimizing active power losses. By optimizing capacitor placement and sizing across the distribution network, the CSA helps prevent overloading specific parts of the network, thereby reducing active power losses.

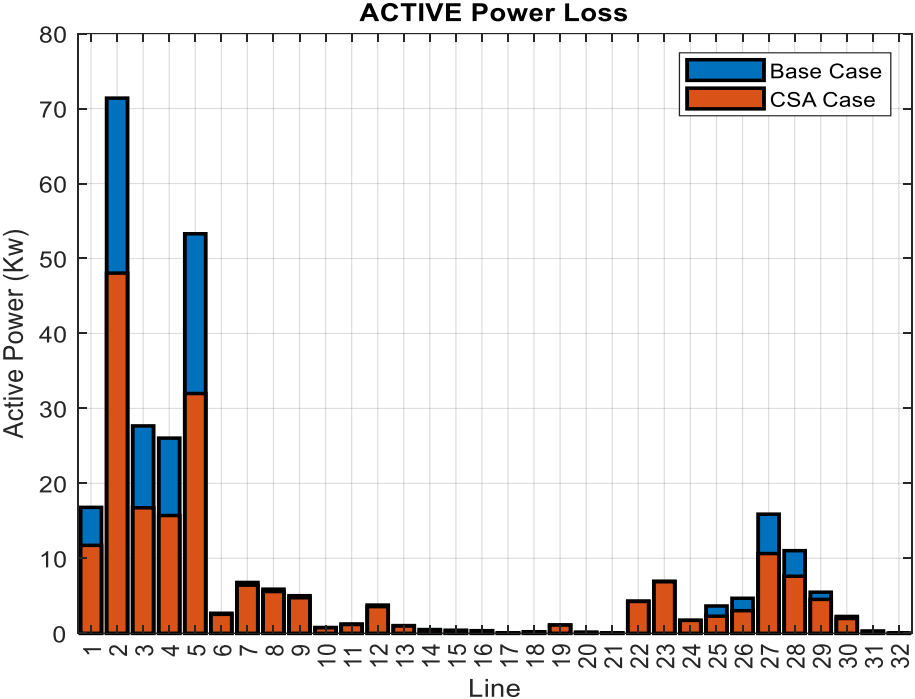

Figure 9: Comparison of CSA and the base case active power losses

The use of the CSA in the sizing and placement of capacitors resulted in a substantial improvement in reactive power loss, as depicted in Figure 10. This improvement is due to the CSA's effective optimization of capacitor placement and sizing, which enhances load balancing, optimal network reconfiguration, and strategic reactive power management. By improving voltage profiles, and optimally placing capacitors, the CSA significantly enhances the overall efficiency of the distribution network, leading to lower reactive power losses and a more stable power supply. By finding the optimal paths for power flow, the CSA reduces the overall reactance of the lines. Shorter and lower reactance paths result in reduced reactive power losses.

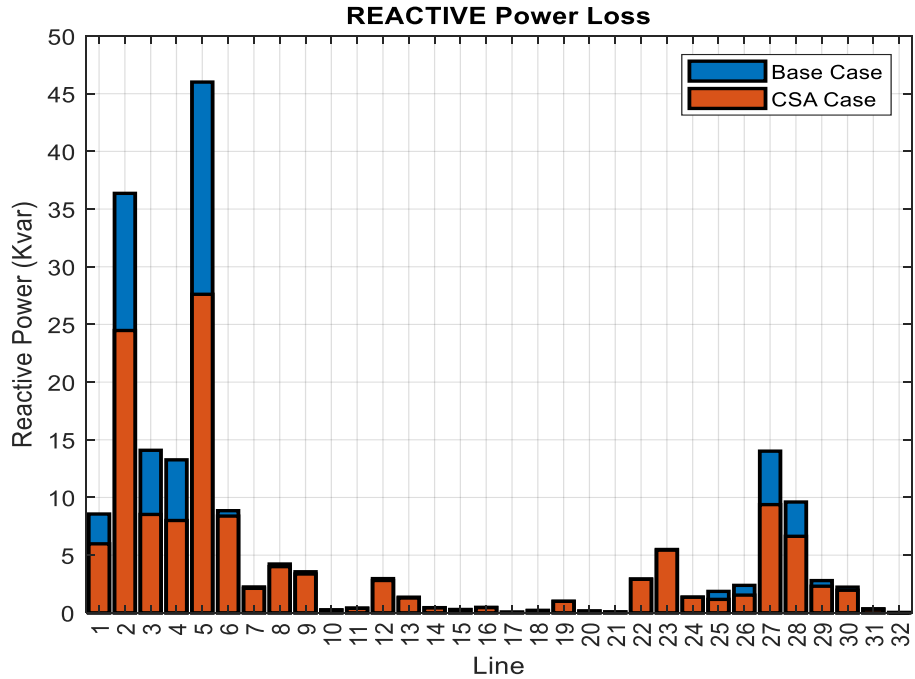

Figure 10: Comparison of CSA and the base case reactive power losses

The reduction in per unit cost from 3.8677 to 3.841, as illustrated in Figure 11, is a result of the CSA's ability to optimize the distribution network effectively. By reducing both active and reactive power losses, improving the efficiency of power flow, and enhancing load management, the CSA achieves significant cost savings. These improvements lead to lower operational costs, reduced energy purchase requirements, and enhanced reliability of the network, culminating in a substantial reduction in per unit cost.

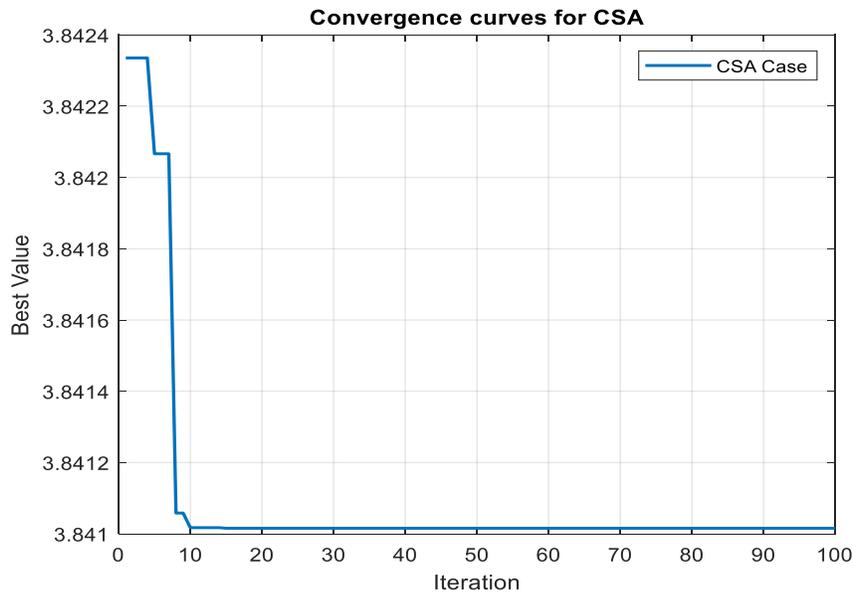

Figure 11: Per unit cost value for CSA case

## 5. Data Validation

The aim of this paper was to explore the optimization performance of the RDN through capacitor sizing and placement based on CSA. In doing so comparison study was conducted with PSO in enhancement of the voltage profile and reduction of network losses. From the voltage profile in Figure 12, there is a significant improvement in voltage profile when CSA and PSO were used as compared with the base case. For instance, in bus 18, the voltage improves from 0.8820 to 0.9021 p.u and 0.90798 p.u, for the base case, PSO and CSA respectively.

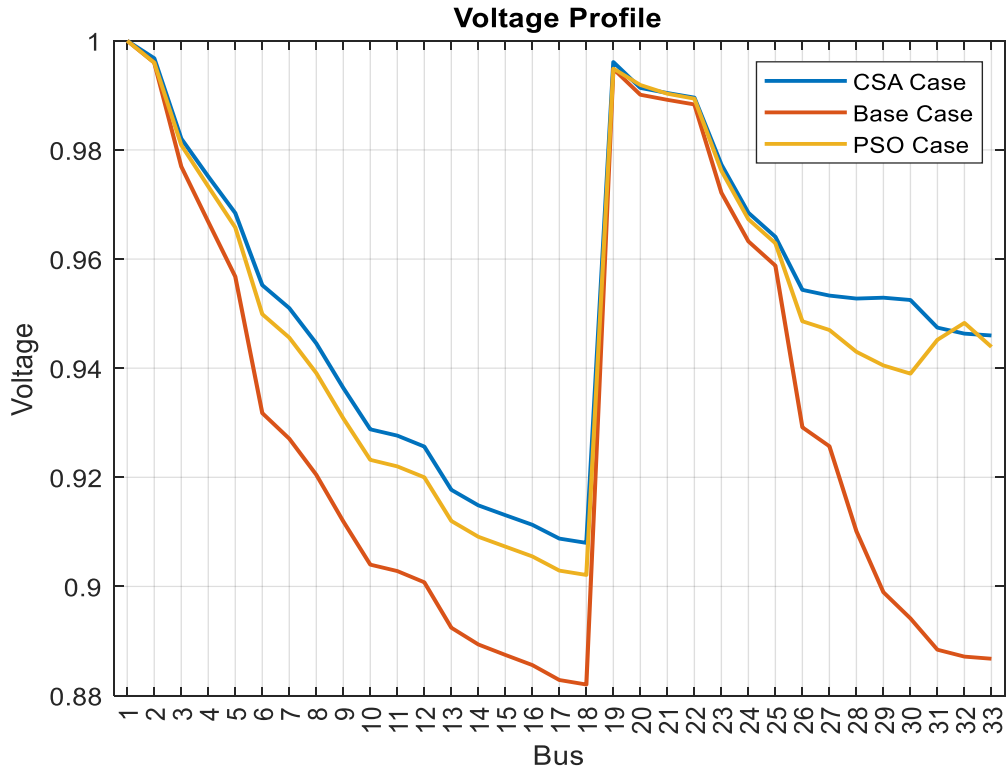

Figure 12: Voltage profile for base, PSO and CSA case

Figure 13 illustrates the proof that the CSA algorithm showed a significant improvement from 71.3728 kW in the base case as compared to 48.0506 kW and 50.7079 kW for PSO and CSA in bus 2. In bus 5 the base case was 53.2985 kW compared with 31.9877 kW to 32.6078 kW for the PSO and CSA.

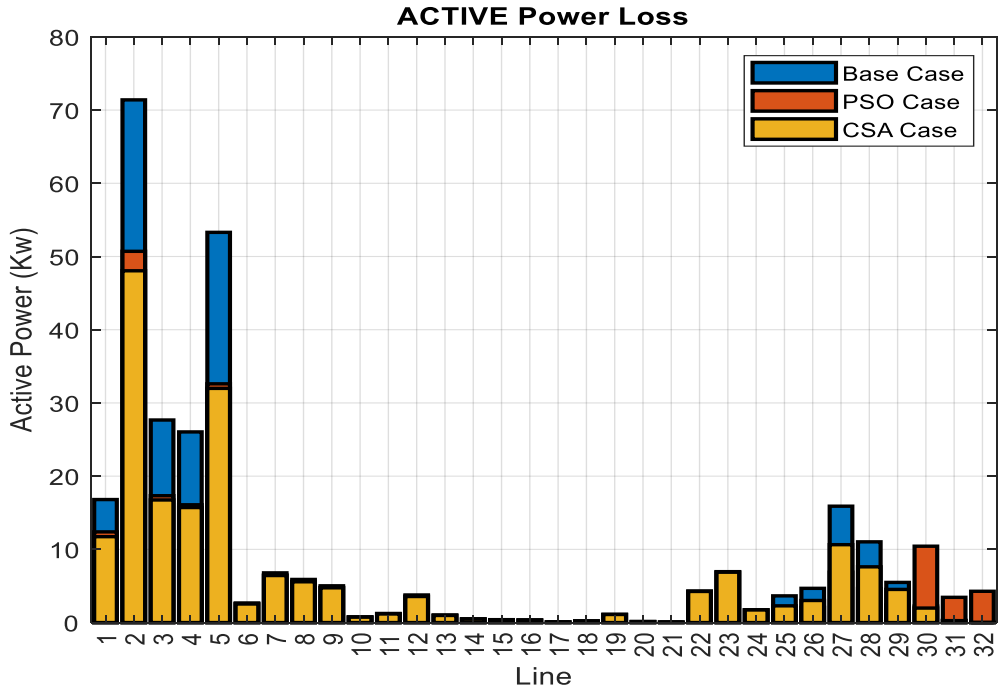

Figure 13: Active power loss for Base, PSO and CSA case

Figure 14 illustrates the affirmation that the reactive power loss from the CSA algorithm showed a significant improvement from 36.3625 Kvar to 25.8271 Kvar for PSO and 24.4736 Kvar for CSA in bus 2 and 46.0098 Kvar for base case compared to 28.1486 Kvar for PSO and 27.6133 Kvar for CSA in bus 5

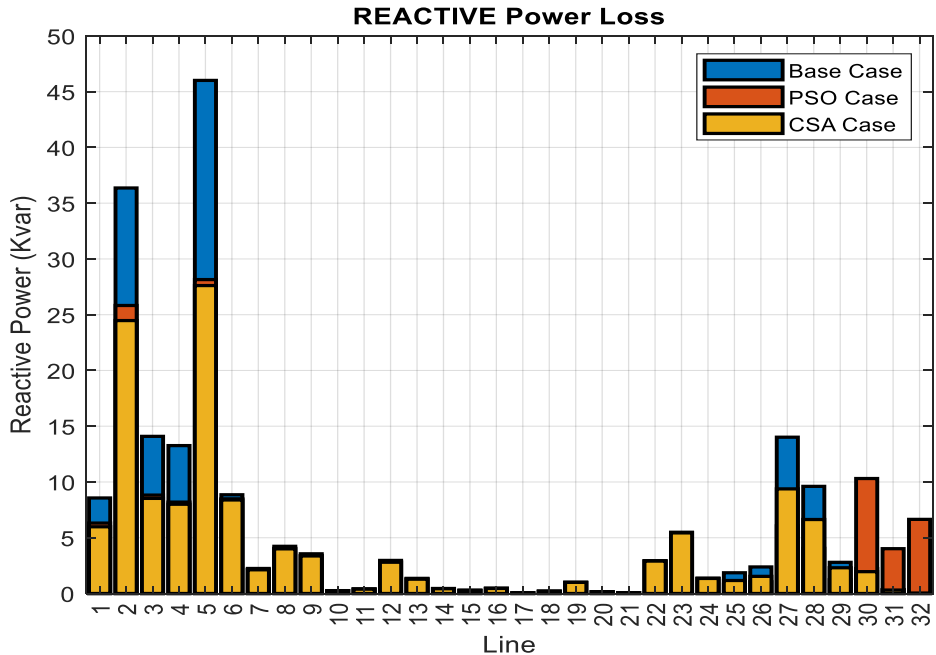

Figure 14: Reactive power loss for Base, PSO and CSA case

Figure 15 illustrates the per unit cost reduction from 4.0452 p.u. to 3.841 p.u using both the PSO and CSA. This significant improvement in cost is due to the algorithms' ability to optimize the distribution network, focusing on reducing both active and reactive power losses. These optimizations result in substantial cost savings, showcasing the exceptional performance of PSO and CSA in enhancing power efficiency and reducing operational costs.

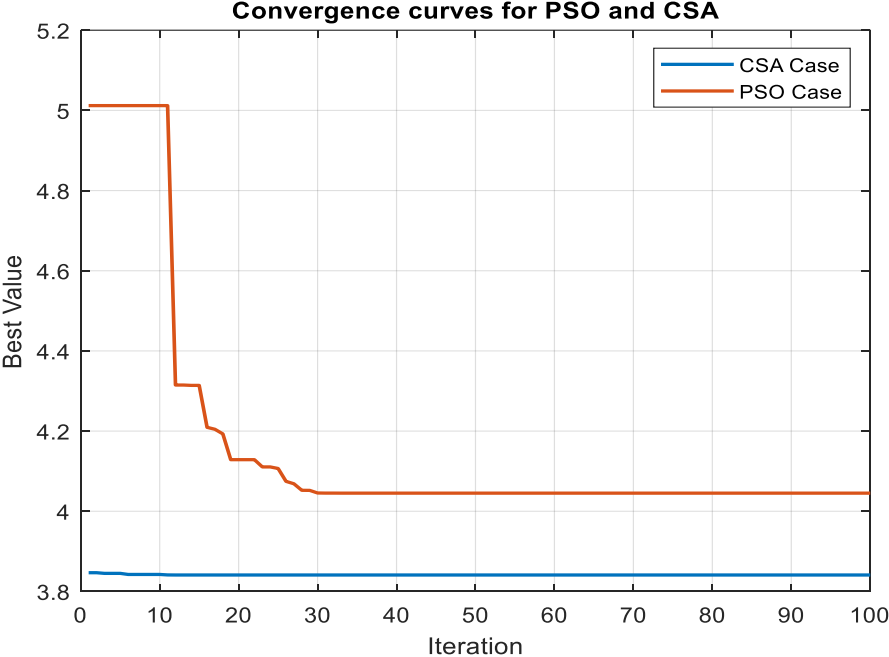

Figure 15: Per unit convergence curve for PSO and CSA

In summary, the CSA algorithm demonstrated exceptional performance in terms of both loss reduction and voltage improvement, as evidenced by the findings presented in Table 1. These outcomes were derived from simulations conducted on a standard IEEE 33 bus system operating at 11kV, comparing the CSA with various other algorithms. The findings unequivocally affirm the effectiveness of the CSA in optimizing power distribution systems.

From Table 1 it is evident that, certain optimization algorithms strategically position capacitors, often at bus 33, aiming to enhance voltage stability, compensate for reactive power demand, optimize load distribution, align with optimization objectives, adhere to system constraints, and reflect algorithm design choices. By leveraging the advantages of this placement, algorithms aim to enhance overall network performance and achieve their optimization goals effectively.

By optimizing both the sizing and placement of capacitors, an array was used to determine the optimal configuration for a single unit. This approach allows the algorithm to enhance the overall efficiency and stability of the power distribution network.

Table 1: Performance comparison of different Algorithms

| Case | Ploss (Kw) | Qloss (Kvar) | Voltage Deviation (VD) | Voltage Stability Index (VSI) | Cap size (Mvar) | Cap location Bus | Best cost (p.u) |
|---|---|---|---|---|---|---|---|
| Base | 281.58 | 187.96 | 2.31 | 0.61 | - | - | - |
| CSA (Crow Search Algorithm) | 195.96 | 132.84 | 1.55 | 0.68 | 1.579 | 30 | 3.841 |
| PSO (Particle Swarm Optimization) [48] | 206.66 | 146.77 | 1.68 | 0.66 | 1.2134 | 33 | 4.0452 |
| ABC (Artificial bee colony) [47] | 206.66 | 146.77 | 1.68 | 0.66 | 1.2134 | 33 | 4.0452 |
| CA (Cultural Algorithm) [45] | 281.58 | 187.96 | 2.31 | 0.61 | 1.2821 | 10 | 5.0119 |
| FA (Firefly Algorithm) [44,46] | 281.58 | 187.96 | 2.31 | 0.61 | 2.0275 | 26 | 5.0119 |
| GA (Genetic Algorithm) [49] | 206.66 | 146.77 | 1.68 | 0.66 | 1.2134 | 33 | 4.0452 |
| IWO (Invasive Weed Optimization) [43] | 211.41 | 153.75 | 1.56 | 0.67 | 1.422 | 33 | 4.0699 |
| TLBO (Teacher Learner Based Optimization) [42] | 206.66 | 146.77 | 1.68 | 0.66 | 1.2134 | 33 | 4.0452 |

## 6.0 Conclusion

This study focused on optimal capacitor sizing and placement in a Radial Distribution Network based on CSA. A standard IEEE 33 bus, 11 kV bus model was modelled in Matlab platform. The base case presented an overall loss of 281.58 kW and 187.96 Kvar upon which after implementing the CSA algorithm the losses were reduced to 195.96 kW and 132.84 Kvar for a 33 bus RDN. This indicated a significate drop of 30.41% and 29.33%.

To validate the efficacy of the CSA algorithm, a series of optimization techniques including Particle Swarm Optimization, Artificial Bee Colony, Cultural Algorithm, Firefly Algorithm, Genetic Algorithm, Invasive Weed Optimization, and Teacher-Learner Based Optimization, were employed as evidenced in Table 1. The results consistently demonstrated the exceptional performance of the algorithm, affirming its applicability across various network topologies, particularly radial distribution networks. Notably, the algorithm significantly enhanced the voltage profile following capacitor sizing and placement within the radial distribution network, indicating its potential for substantial loss reduction. This study underscores the adaptability of the algorithm for utility companies seeking to optimize their networks, irrespective of topology, thus maximizing profitability.

However, it's important to note that this study exclusively focused on radial distribution networks. Further investigation into other network topologies, such as ring and interconnected networks, could provide valuable insights and opportunities for additional research and optimization.

**Acknowledgment**

This research was supported by Dedan Kimathi University of Technology in the form of a postgraduate student research funding.

**Declaration of Competing Interest**

The authors declare that they have no known competing financial interests or personal relationships that could have appeared to influence the work reported in this paper.

**Data availability Section**

Data included in this article has been referenced and the rest are available on request.